\documentclass[a4paper,fleqn,usenatbib]{mnras}

\usepackage[T1]{fontenc}
\usepackage{ae,aecompl}

\usepackage{graphicx}	% Including figure files
\usepackage{amsmath}	% Advanced maths commands
\usepackage{amssymb}	% Extra maths symbols

\title[Milky Way star clusters]{Nearly coeval intermediate-age Milky Way star clusters
at very different dynamics evolutionary stages}

\author[Piatti, Angelo \& Dias]{
Andr\'es E. Piatti$^{1,2}$\thanks{E-mail: andres.piatti@unc.edu.ar}, Mateus S. Angelo$^{3}$ and Wilton S. Dias$^{4}$\\
$^{1}$Consejo Nacional de Investigaciones Cient\'{\i}ficas y T\'ecnicas, Godoy Cruz 2290, C1425FQB, 
Buenos Aires, Argentina\\
$^{2}$Observatorio Astron\'omico de C\'ordoba, Laprida 854, 5000, 
C\'ordoba, Argentina\\
$^{3}$Centro Federal de Educa\c{c}\~ao Tecnol\'ogica de Minas Gerais, Av. Monsenhor Luiz de Gonzaga, 103, 37250-000 Nepomuceno, MG, Brazil\\
$^{4}$UNIFEI, Instituto de F\' {\i}sica e Qu\' {\i}mica, Universidade Federal de Itajub\'a, Av. BPS 1303 Pinheirinho, 37500-903 Itajub\'a, MG, Brazil \\
}

\date{Accepted XXX. Received YYY; in original form ZZZ}

\pubyear{2019}

\begin{document}
\label{firstpage}
\pagerange{\pageref{firstpage}--\pageref{lastpage}}
\maketitle

% Abstract of the paper
\begin{abstract}
We report astrophysical properties of 12 Milky Way open clusters located
beyond a 2 kpc circle around the Sun by using deep optical photometry.
We estimated their age and metallicities on the basis of a maximum likelihood
approach using astrometric members determined from Gaia DR2 data.
The studied clusters turned out to be
of intermediate-age (0.8 - 4.0 Gyr), with metallicities spanning the range
[Fe/H] $\sim$ -0.5 - +0.1 dex, and distributed within the general
observed trend of the Milky Way disc radial and perpendicular metallicity
gradients. As far as we are aware, these are the first 
metal abundance estimates derived for these clusters so far.
From the constructed stellar density radial profiles and cluster mass
functions we obtained a variety of structural and internal dynamics
evolution parameters. They show that while the innermost cluster regions
would seem to be mainly shaped according to the respective internal dynamics 
evolutionary stages, the outermost ones would seem to be slightly more 
sensitive to the Milky Way tidal field. The nearly coeval studied clusters 
are experiencing different levels of two-body relaxation following 
star evaporation; those at more advanced stages being more compact objects.
Likewise, we found that the more important the Milky way tides, the
larger the Jacobi volume occupied by the clusters,  irrespective of their
actual sizes and internal dynamics evolutionary stages.

\end{abstract} 

% Select between one and six entries from the list of approved keywords.
% Don't make up new ones.

%%%%%%%%%%%%%%%%%%%%%%%%%%%%%%%%%%%%%%%%%%%%%%%%%%

%%%%%%%%%%%%%%%%% BODY OF PAPER %%%%%%%%%%%%%%%%%

\section{Introduction}

The study of Galactic open clusters has long helped improve our understanding 
of the Milky Way disc formation and evolution. For instance,
from their positions, ages and metallicities, the radial metallicty gradient as well
as that perpendicular to the Galactic plane  have been derived, which in turn have
been used to constrain Milky Way formation theoretical models 
\citep{metal09,sk2018}.
Different photometric and spectroscopic surveys have been exploited in order to
derive improved cluster parameters and first estimates of the astrophysical
properties of unstudied ones \citep[e.g.][]{cantatgaudinetal2018,carreraetal2019}. 
Although there has been an interesting progress in constructing
open clusters' parameter catalogues \citep{detal02,kharchenkoetal2013}, 
the remaining work is still huge, because of the growing number of identified
new open clusters \citep{cantatgaudinetal2018,castroginardetal2018,ferreiraetal2019}.

With the aim of contributing to a comprehensive knowledge of the open cluster 
system, we searched the 
National Optical Astronomy Observatory (NOAO) Science Data Management 
Archives\footnote{http://www.noao.edu/sdm/archives.php.} looking for Washington 
photometric system images centred on mostly unstudied open cluster fields. We chose
the Washington photometric system because of our experience in estimating
star cluster fundamental parameters and its ability in estimating cluster metallcities
\citep[see, e.g.][and references therein]{petal04,piattietal2017a}. From the 
search, we found that the Cerro Tololo Interamerican Observatory (CTIO) programme
no. 2008A-0001 (PI: Clari\'a) was aimed at observing nearly 80 mostly
unstudied open clusters. We have started to analyse them in 
\citet{angeloetal2018} and \citet{angeloetal2019b}. Here, we focus on all the
remaining clusters with ages $\sim$ 1 Gyr, because their structural
properties tell us about the wide range of internal dynamical evolutionary
stages they can span.

In Section 2 we present the unpublished publicly available Washington data
sets used in this work, alongside a brief description of the procedure followed
to obtained the standardised photometry, and hence, the resulting clusters'
features. Section 3 deals with the analysis of their astrophysical properties
in the context of the Milky Way chemical evolution framework, and of their
structural parameters. The latter are discussed to the light of the disruption
processes due to stellar evolution and tidal forces. Finally, Section 4
summarises the main conclusions of this work.

\section{cluster parameter estimates}

The collected $C$ Washington and $R$ Kron-Cousins images were obtained at the 
CTIO 0.9m telescope with the Tek2K CCD imager attached (scale = 0.4 arcsec 
pixel$^{\rm -1}$, FOV = 13.6$\times$13.6 arcmin$^{\rm 2}$). We downloaded
calibrations frames (bias, dome- and sky-flats), standard star field and
programme images (see Table~\ref{tab:table1}), which were processed following 
the standard pipeline with
{\sc quadred} tasks in the {\sc iraf}\footnote{IRAF is distributed by the 
National Optical Astronomy Observatories, which is operated by the Association 
of Universities for Research in Astronomy, Inc., under contract with the National
Science Foundation.} package. PSF photometry was obtained through the
{\sc starfinder} code \citep{diolaitietal2000} by modelling the
PSF from high signal-to-noise and relatively isolated stars and by keeping only
magnitudes of stars with the correlation coefficients between the measured
profile and the modelled PSF greater than 0.7. Astrometric coordinates for
all the stars were obtained by transforming the CCD ones into those given by {\it Gaia}
DR2 \citep{gaiaetal2016,gaiaetal2018b}, while their magnitudes in the standard $CT_1$ 
Washington system were obtained from transforming the instrumental $c,r$ 
magnitudes using the transformation equations derived in \citet{angeloetal2018}.

\begin{table*}
\caption{Observations log of the studied open clusters.}
\label{tab:table1}
\begin{tabular}{@{}lccccccc}\hline
Cluster & $\rmn{RA}$     & $\rmn{DEC}$     & $\ell$     & $b$     & Filter     & Exposure  & FWHM  \\   
            &  ($\rmn{h}$:$\rmn{m}$:$\rmn{s}$) & ($\degr$:$\arcmin$:$\arcsec$) & ($^{\circ}$) & ($^{\circ}$) &   
& (s) & ($\arcsec$)\\\hline            

ESO\,96-SC04   & 13:15:16.0 &-65:55:16 & 305.36 &-03.16 & $C$ & 2$\times$30, 2$\times$300 & 1.1,1.1,1.1,1.1\\
               &            &          &        &        & $R$ & 2$\times$5,2$\times$30    & 1.0,1.0,1.0,1.0\\
ESO\,137-SC23  & 16:24:30.5 &-61:43:59 & 325.50 &-08.59 & $C$ & 2$\times$30,45,300,450    & 1.0,1.0,1.0,1.1,1.1\\
               &            &          &        &        & $R$ & 5,7,30,45                 & 1.0,1.0,1.0,1.0\\
ESO\,334-SC02  & 17:36:56.9 &-42:14:06 & 347.75 &-05.51 & $C$ & 2$\times$30,2$\times$300  & 1.2,1.2,1.2,1.2\\      
               &            &          &        &        & $R$ & 3$\times$5,2$\times$30    & 1.1,1.1,1.1,1.1,1.0\\
ESO\,371-SC25  & 08:53:00.0 &-35:28:01 & 258.00 &+05.86 & $C$ & 15,20,2$\times$350        & 1.1,1.1,1.1,1.1\\
               &            &          &        &        & $R$ & 15,30,60,2$\times$120     & 1.0,1.0,1.1,1.0,1.0\\
ESO\,429-SC13  & 07:41:03.0 &-30:44:06 & 245.63 &-03.94 & $C$ & 15,3$\times$30,90         & 1.1,1.2,1.2,1.2,1.2\\
               &            &          &        &        & $R$ & 2$\times$10,60            & 1.2,1.2,1.2\\
NGC\,4230      & 12:17:20.4 &-55:07:12 & 298.02 &+07.42 & $C$ & 30,45                     & 1.3,1.3\\
               &            &          &        &        & $R$ & 5,7                       & 1.2,1.2\\
NGC\,4337      & 12:24:00.0 &-58:07:01 & 299.30 &+04.56 & $C$ & 2$\times$30, 2$\times$300 & 1.1,1.1,1.1,1.1\\
               &            &          &        &        & $R$ & 2$\times$5,2$\times$30    & 1.0,1.0,1.1,1.1\\
NGC\,6525      & 18:02:00.0 &+11:01:24 & 037.36 &+15.91 & $C$ & 10,15,2$\times$300        & 1.2,1.2,1.2,1.2\\
               &            &          &        &        & $R$ & 2$\times$5,2$\times$30    & 1.1,1.1,1.1,1.1\\
Ruprecht\,37   & 07:49:51.6 &-17:13:30 & 234.92 &+04.54 & $C$ & 2$\times$40,300,450       & 1.0,1.0,1.1,1.1\\
               &            &          &        &        & $R$ & 10,3$\times$30,90         & 1.0,1.0,1.0,1.0,1.1\\
Ruprecht\,74   & 09:21:00.0 &-37:07:01 & 263.03 &+08.96 & $C$ & 30,60,2$\times$600        & 1.3,1.3,1.3,1.3\\
               &            &          &        &        & $R$ & 10,15,2$\times$120        & 1.2,1.2,1.2,1.2\\
Ruprecht\,104  & 12:25:01.2 &-60:26:42 & 299.67 &+02.25 & $C$ & 60,90,900                 & 1.2,1.2,1.3\\
               &            &          &        &        & $R$ & 2$\times$30,180           & 1.2,1.2,1.2\\
Ruprecht\,163  & 11:04:53.3 &-67:56:35 & 293.16 &-07.11 & $C$ & 90,180,900                & 1.1,1.1,1.2\\
                &            &          &       &        & $R$ & 40,60,2$\times$180        & 1.1,1.1,1.1,1.1\\\hline
\end{tabular}
\end{table*}

\subsection{Selection of cluster members}

In order to disentangle probable cluster members from field stars in
the present Washington data set, we applied the method developed
by \citet{angeloetal2019a}, which runs on the basis of {\it Gaia} DR2
proper motions ($\mu$$_{\alpha}$, $\mu$$_{\delta}$) and parallaxes ($\varpi$).
The method compares in the 3D parameter space ($\mu$$_{\alpha}$, 
$\mu$$_{\delta}$, $\varpi$)
the distributions of stars  located within the cluster tidal radius and beyond that. Such 
a comparison is carried out in three steps,  namely: a uniform grid of cells is
built with sizes  $\sim$ 10$\times$$\Delta (\mu$$_{\alpha})$, 
10$\times$$\Delta (\mu$$_{\delta})$
and 1$\times$$\Delta (\varpi)$, where $\Delta (\mu$$_{\alpha})$,
$\Delta (\mu$$_{\delta})$ and $\Delta (\varpi)$ are the mean uncertainties
for the whole sample of cluster and control field stars. Then, the 3D Gaussian distributions 
of stars located
inside and outside the cluster tidal radius for each defined cell are compared, 
looking for local stellar overdensities that are statistically distinguishable from 
the distribution of field stars, and flagged
those stars located inside the tidal radius with 1 (probable member) or 0 (non-member).
Finally, the individual membership likelihood for stars flagged  1 are estimated
from the comparison of the number of them in a particular cell with respect to the
weighted average number of them throughout all the cells. 

\begin{figure*}
\includegraphics[width=\textwidth]{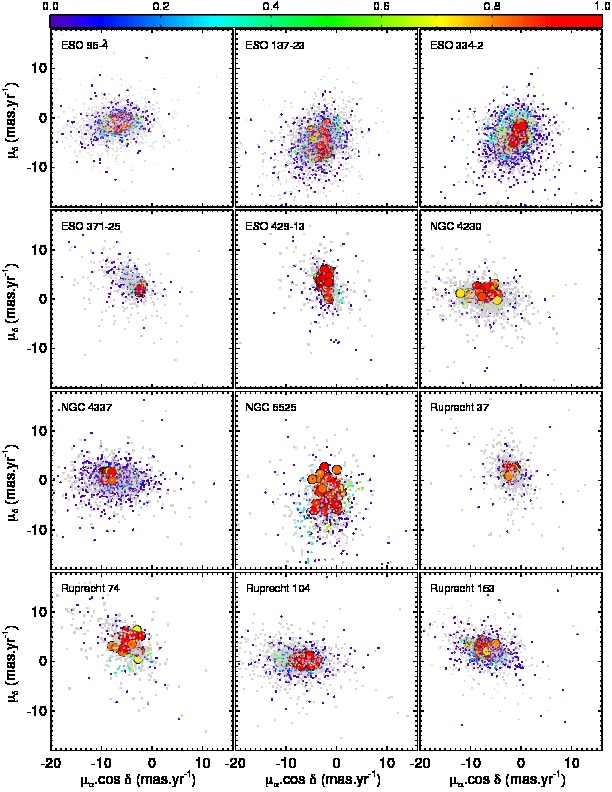}
\caption{Vector-point diagrams for stars measured in the studied cluster fields.
Large and small symbols represents probable members and non-members,
respectively. In particular, small grey symbols correspond to field stars
located beyond the tidal radius. Top colour-coded bar depicts the membership probabilities.}
\label{fig:fig1}
\end{figure*}

\begin{figure*}
\includegraphics[width=\textwidth]{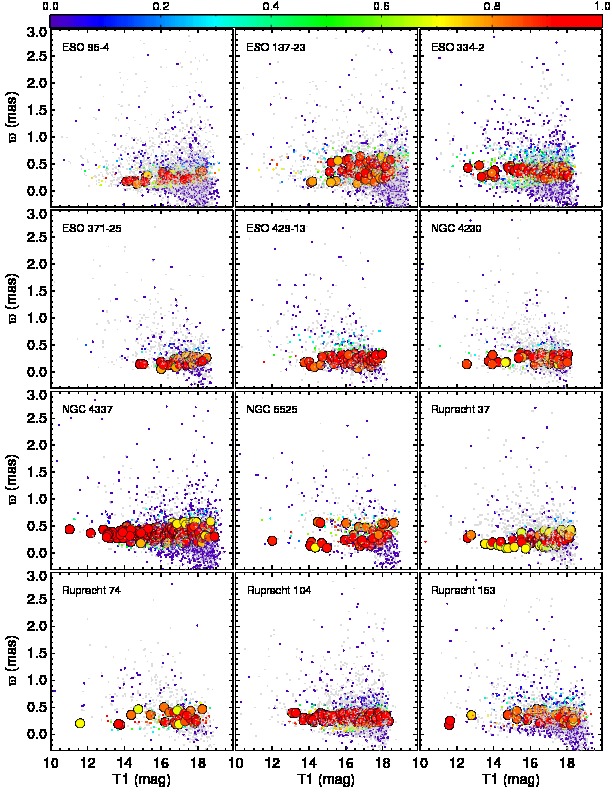}
\caption{Parallaxes ($\varpi$) versus $T_1$ magnitudes diagram for stars measured in the studied cluster fields.
Symbols and colours are as in Fig.~\ref{fig:fig1}. Top colour-coded bar depicts the membership probabilities. }
\label{fig:fig2}
\end{figure*}

We repeated these steps for cells one third smaller and bigger than those initially
used, so that for each star we assigned 27 different membership likelihood, whose
median was adopted as the final membership probability ($P$). In the subsequent
analysis we considered probable clusters members those with $P \ge$ 0.7.
Figs.~\ref{fig:fig1} and \ref{fig:fig2} show, respectively, the vector-point diagrams and the
$\varpi$ versus $T_1$ plane for all the stars measured in the field of the
selected clusters coloured according to the resulting membership probabilities
(colour-coded bar placed at the top of the figures). 
Large and small symbols represent probable members and
non-members, respectively. Particularly small grey symbols are field stars 
located beyond the cluster tidal radii.

\subsection{Cluster fundamental properties}

Mean ages (log($t$ /yr)), metallicities ([Fe/H]), true distances moduli ($(m-M)_o$) 
and reddenings ($E(B-V)$) for the cluster sample were estimated using stars with $P \ge$ 0.7.  
In some cases, we added some few stars without {\it Gaia} DR2 data to which we
assigned photometric membership probabilities higher than 0.7 on the basis of the
procedure developed by \citet{maiaetal2010}. Basically, their method compares the
distributions of stars in the $T_1$ versus $C-T_1$ colour-magnitude diagram (CMD) located
inside and outside the cluster tidal radius. Such distributions are built
using a grid of cells uniformly distributed in the CMD, and the comparison iterated for
grids with cells of different sizes and then averaged all the
individual photometric memberships.

Since the four cluster properties are not fully independent one to each other,
we employed the Automated Stellar Cluster Analysis suit of functions 
\citep[ASteCA,][]{pvp15} -- devised for running for Washington photometry -- to build
thousands of synthetic cluster CMDs which
were matched to the observed ones in order to find the best solutions for the
cluster astrophysical parameters \citep[see, e.g.][]{p17b,pc2017}. These synthetic CMDs 
are generated using the theoretical isochrones of \citet{betal12}, an initial
mass function according to \citet{Chabrier_2001}, a 50 per cent of binaries 
\citep{von_Hippel_2005} with secondary masses drawn
from a uniform distribution between the mass of the primary star
and a fraction of 0.7 of it, and the appropriate magnitude completeness
and photometric errors. We used a grid of $E(B-V)$, $(m-M)_o$, log($t$ /yr) and [Fe/H]
values with steps of $\Delta$($E(B-V)$) = 0.01 mag, $\Delta$($(m-M)_o$) = 0.1 mag, 
$\Delta$(log($t$ /yr)) = 0.01, and $\Delta$($Z$) = 0.002 ([Fe/H] = log($Z/Z_\odot$)
with $Z_\odot$ = 0.0152, \citet{betal12}), respectively. Particularly, a first
guess for the distance moduli were obtained from $(m-M)_o$ = 5$\times$ log(100/$\varpi$),
where $\varpi$ are the mean cluster parallaxes derived from astrometric membership.

Fig.~\ref{fig:fig3} shows the $T_1$ versus $C-T_1$ CMDs for all the
measured stars in the fields of the studied clusters. We have highlighted those
with $P \ge$ 0.7 and superimposed the isochrones corresponding the best-matched
synthetic cluster CMDs, while Table~\ref{tab:table2} lists the resulting
cluster astrophysical properties.

\begin{table*}
\caption{Astrophysical parameters of the studied open clusters. Structural parameters, namely,
core ($r_c$), half-ligt ($r_h$), tidal ($r_t$) and Jacobi ($r_J$) radii are also listed.}
\label{tab:table2}
\begin{tabular}{@{}lcccccccccc}\hline
Cluster        &  $E(B-V)$ & $(m-M)_o$ & log($t$ /yr) & [Fe/H] & $r_c$ & $r_h$ & $r_t$ & $r_J$ & $M_{cls}$ & $t_{rh}$ \\
               &   (mag)   &  (mag)    &              & (dex)  &  (pc) &  (pc) &  (pc) &  (pc) & ($M_\odot$) & (Myr)  \\\hline          
ESO\,96-SC04   & 0.63$\pm$0.10 & 13.2$\pm$0.5 & 8.90$\pm$0.20 & 0.05$\pm$0.15& 0.8$\pm$0.3 & 0.9$\pm$0.1 & 1.9$\pm$0.3 & 4.0$\pm$0.5 & 73$\pm$11 & 3.0$\pm$0.7 \\
ESO\,137-SC23  & 0.40$\pm$0.10 & 12.1$\pm$0.3 & 9.45$\pm$0.15 &-0.13$\pm$0.23& 0.5$\pm$0.1 & 0.8$\pm$0.1 & 1.9$\pm$0.7 & 3.7$\pm$0.3 & 71$\pm$8  & 3.5$\pm$0.8 \\
ESO\,334-SC02  & 0.65$\pm$0.05 & 11.9$\pm$0.2 & 8.95$\pm$0.10 & 0.10$\pm$0.13& 0.9$\pm$0.2 & 1.4$\pm$0.2 & 3.6$\pm$1.3 & 4.6$\pm$0.4 &161$\pm$15 & 6.2$\pm$1.7 \\
ESO\,371-SC25  & 0.55$\pm$0.10 & 13.4$\pm$0.3 & 9.20$\pm$0.10 &-0.21$\pm$0.18& 1.2$\pm$0.2 & 1.7$\pm$0.2 & 3.6$\pm$0.9 & 5.8$\pm$0.4 & 58$\pm$9  & 8.0$\pm$1.9 \\
ESO\,429-SC13  & 0.48$\pm$0.05 & 13.0$\pm$0.3 & 8.95$\pm$0.15 &-0.13$\pm$0.15& 1.8$\pm$0.5 & 2.5$\pm$0.4 & 4.9$\pm$1.7 & 7.3$\pm$0.4 &109$\pm$13 &13.8$\pm$3.6 \\
NGC\,4230      & 0.45$\pm$0.15 & 12.7$\pm$0.5 & 9.05$\pm$0.20 &-0.06$\pm$0.19& 1.3$\pm$0.3 & 1.6$\pm$0.2 & 3.1$\pm$0.7 & 4.4$\pm$0.4 & 74$\pm$10 & 7.2$\pm$1.7 \\
NGC\,4337      & 0.62$\pm$0.05 & 11.9$\pm$0.3 & 8.95$\pm$0.10 &-0.46$\pm$0.33& 1.6$\pm$0.2 & 2.5$\pm$0.2 & 5.8$\pm$1.0 & 8.4$\pm$0.6 &503$\pm$28 &20.1$\pm$3.2 \\
NGC\,6525      & 0.45$\pm$0.10 & 12.6$\pm$0.4 & 9.10$\pm$0.15 &-0.32$\pm$0.24& 0.4$\pm$0.2 & 0.7$\pm$0.2 & 2.4$\pm$1.1 & 3.1$\pm$0.3 & 44$\pm$7  & 2.2$\pm$1.1 \\
Ruprecht\,37   & 0.16$\pm$0.05 & 12.5$\pm$0.2 & 9.65$\pm$0.10 &-0.32$\pm$0.24& 1.2$\pm$0.2 & 1.3$\pm$0.1 & 2.1$\pm$0.2 & 6.3$\pm$0.4 & 75$\pm$8  & 7.7$\pm$1.5 \\
Ruprecht\,74   & 0.38$\pm$0.10 & 12.4$\pm$0.4 & 9.00$\pm$0.20 & 0.00$\pm$0.22& 0.5$\pm$0.1 & 0.8$\pm$0.1 & 1.7$\pm$0.4 & 4.4$\pm$0.3 & 38$\pm$7  & 2.2$\pm$0.4 \\
Ruprecht\,104  & 0.63$\pm$0.10 & 12.3$\pm$0.4 & 8.90$\pm$0.10 & 0.00$\pm$0.17& 2.3$\pm$0.6 & 3.1$\pm$0.5 & 5.2$\pm$1.2 & 6.2$\pm$0.5 &214$\pm$18 &20.6$\pm$5.3 \\
Ruprecht\,163  & 0.28$\pm$0.15 & 12.4$\pm$0.4 & 9.40$\pm$0.15 &-0.13$\pm$0.23& 0.3$\pm$0.1 & 0.6$\pm$0.1 & 1.8$\pm$0.6 & 3.8$\pm$0.3 & 42$\pm$7  & 1.9$\pm$0.4 \\\hline
\end{tabular}
\end{table*}

\begin{figure*}
\includegraphics[width=0.95\textwidth]{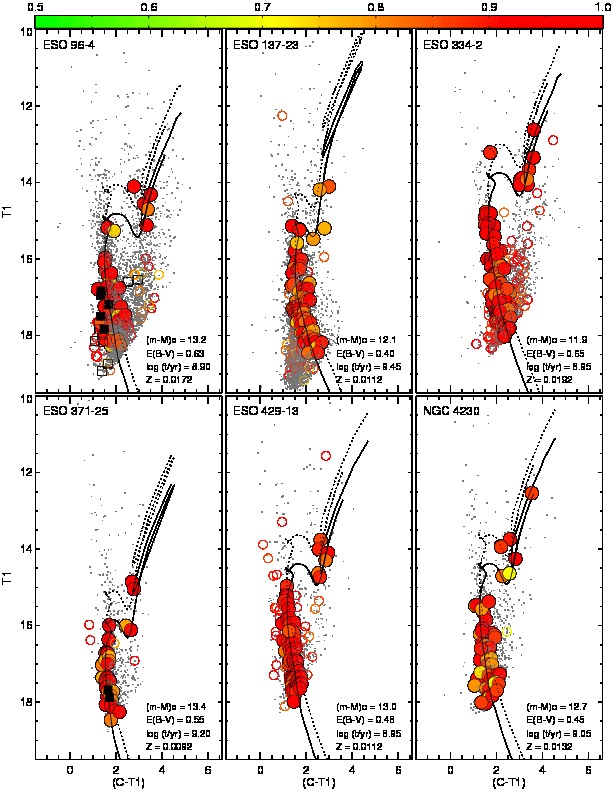}
\caption{$T_1$ versus $C-T_1$ diagrams for stars measured in the studied cluster fields.
Large filled and open circles represent probable members and non-members,
respectively.  In particular, small grey symbols correspond to field stars
located beyond the tidal radius. Top colour-coded bar depicts the membership probabilities. Black filled and open squares are stars without 
{\it Gaia} DR2 data and with photometric membership probabilities higher and lower 
than 0.7, respectively, obtained from the procedures developed by \citet{maiaetal2010}.
The isochrones for the best-matched synthetic cluster CMDs and those shifted by -0.75 
mag in $T_1$ to show loci of unresolved binaries with equal mass components are
superimposed with solid and dotted lines, respectively.}
\label{fig:fig3}
\end{figure*}

\setcounter{figure}{2}
\begin{figure*}
\includegraphics[width=0.95\textwidth]{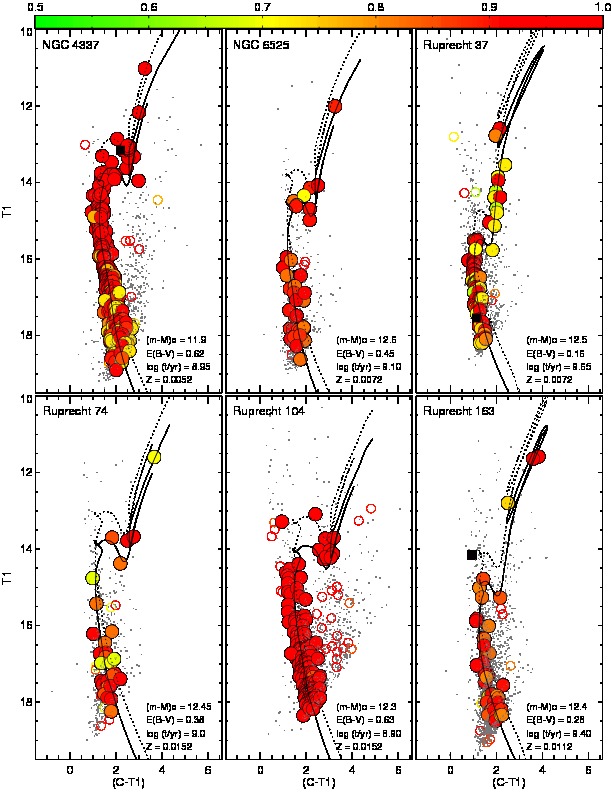}
\caption{continued.}
\label{fig:fig3}
\end{figure*}

\subsection{Cluster structural parameters and masses}

\begin{figure*}
\includegraphics[width=0.95\textwidth]{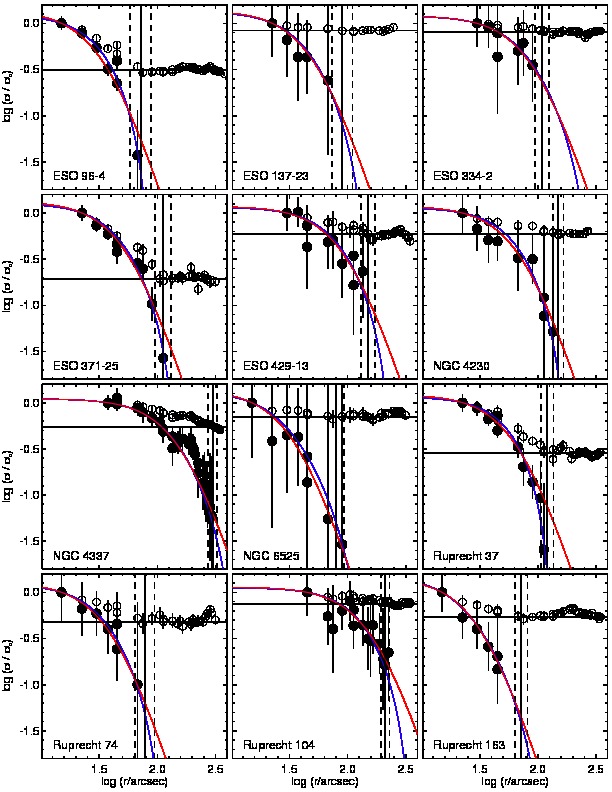}
\caption{Observed and background subtracted stellar density radial profiles for the
studied open clusters drawn with 
open and filled circles, respectively. Error bars are also drawn.
The mean background levels are indicated with
horizontal lines, while the cluster radii with their uncertainties are depicted with vertical
solid and dashed lines, respectively. The fitted \citet{king62}'s and \citet{plummer11}'s
curves are superimposed with blue and red lines, respectively.}
\label{fig:fig4}
\end{figure*}

We used all the measured stars in the selected cluster fields to construct
cluster stellar density radial profiles by counting the number of stars within
concentric rings centred on the clusters' centres. We considered annuli of the
same width, from 0.50 up to 1.50 arcmin, in steps of 0.25 arcmin.
From the five constructed radial profiles per cluster we computed the average
stellar density and dispersion as a function of the distance from the cluster centre
and built the mean cluster radial profiles shown with open circles and error bars in
Fig.~\ref{fig:fig4}. As can be seen, the background level of each cluster has
been properly traced, so that we adopted the mean value of them to be subtracted to the
observed radial profiles and obtain the background subtracted ones (filled circles).
The cluster radii, defined as the distance from the
cluster centre where the combined cluster plus background stellar
density profile is no longer readily distinguished from a constant
background value within 1$\sigma$ of its fluctuation, are also depicted with vertical
lines.

\citet{king62}'s (eq.(1)) and \citet{plummer11}'s (eq.(2)) models were fitted to the 
background subtracted radial profiles in order to derive the cluster core ($r_c$),
half-light ($r_h$) and tidal ($r_t$) radii, respectively. $r_h$ is related to the Plummer 
radius $a$ by the relation $r_h$ $\sim$ 1.3$a$. We used a grid of ($r_c$,$a$,$r_t$) values 
to fit the radial profiles by $\chi^2$ minimisation. Fig.~\ref{fig:fig4} illustrates the 
resulting best-fitted curves.

\begin{equation}
  \sigma(r) \propto \left( \frac{1}{\sqrt{1+(r/r_c)^2}} - \frac{1}{\sqrt{1+(r_t/r_c)^2}} \right)^2,
\end{equation}

\begin{equation}
 \sigma(r) \propto \frac{1}{(1+(r/a)^2)^2}.
\end{equation}

We also derived the cluster Jacobi radii ($r_J$) -- the distance from the cluster centre
beyond which the Milky Way gravitational field dominates the stellar dynamics --
using the expression:

\begin{equation}
r_J = \left(\frac{M_{cls}}{3 M_{MW}}\right)^{1/3}\times R_{GC}
\end{equation}

\noindent where $M_{cls}$ is the cluster mass and $M_{MW}$ is the Milky Way mass comprised
within a radius equal to the Galactocentric cluster distance ($R_{GC}$). The latter is
obtained from the cluster Galactic coordinates ($l$,b) (see Table~\ref{tab:table1}) and
the cluster heliocentric distance $d$ = 10$\times$10$^{(m-M)_o/5}$ (see Table~\ref{tab:table2}).
As for the Milky Way mass, we used $M_{MW}$ $\sim$ 1.0$\times$10$^{11}$$M_\odot$ 
\citep{cc94,bonattoetal2005,tayloretal2016}. In order to obtain the cluster masses, we employed
the individual masses of probable members ($P \ge 0.7$) from interpolation from the corresponding 
theoretical isochrones \citep{betal12}, properly shifted according to the clusters distance modulus and reddening, to build the cluster mass functions, i.e., $\phi$($m$) =
$dN/dm$. For each mass bin, $\phi$($m$) was weighted by the star membership 
probabilities, corrected by photometric completeness \citep[see Fig. 2 of][]{angeloetal2018} 
and assumed Poisson statistics for uncertainty
determination. $M_{cls}$ were then estimated from summing the masses along the
different mass bins of the observed $\phi$($m$) with uncertainties coming from
propagation of errors.

Finally, we estimated the half-light relaxation times from \citep{sh71} :

\begin{equation}
t_h = \frac{8.9\times 10^5 M_{cls}^{1/2} r_h^{3/2}}{\bar{m} log_{10}(0.4M_{cls}/\bar{m})}
\end{equation}

\noindent where $\bar{m}$ is the average mass of the cluster stars obtained from the
cluster mass distribution functions. The resulting different radii ($r_c$, $r_h$, $r_t$, $r_J$),
cluster photometric masses and relaxation times are listed with their uncertainties in 
Table~\ref{tab:table2}.

\section{Analysis and discussion}

\begin{figure*}
\includegraphics[width=\textwidth]{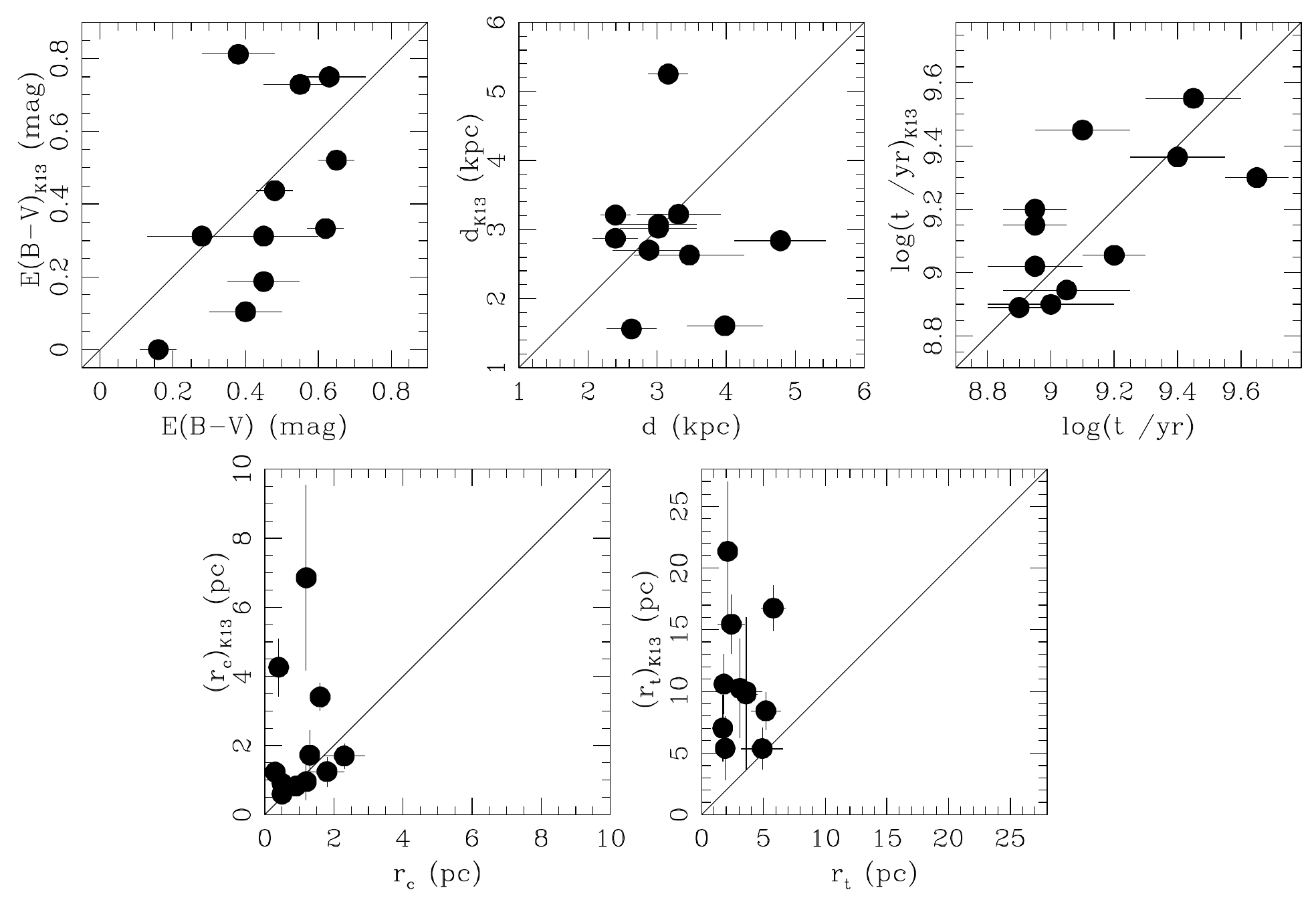}
\caption{Comparison of astrophysical parameters derived by K13
and in this work. Whenever available, error bars are also drawn. The solid line 
represents the identity relationship.}
\label{fig:fig5}
\end{figure*}

We first compared the resulting cluster parameters with those included in the 
\citet[][hereafter K13]{kharchenkoetal2013} open cluster catalogue, which were
derived from the 2MASS database \citep{setal2006}. Note that they cleaned 
cluster CMDs from the contamination of field stars on the basis
of the PPMXL astrometric catalogue \citep{roseretal2010},
thus dealing with the brighter parts of the cluster CMDs
\citep{kharchenkoetal2012}. The results obtained by K13 are based on
photometric data less  deep than the present  Washington photometric data sets
and on assumed solar values for the cluster metallicities of 
those not published in \citet[][version 3.5, January 2016, hereafter  DAML02]{detal02}.
Fig.~\ref{fig:fig5} illustrates the results of the comparison,
where we plotted the values for 11 clusters in common (ESO\,96-SC04 is the only 
one not included in K13). As can be seen, there is not a tight agreement, particularly
for the clusters' tidal radii. 

K13 did not fit the stellar density cluster radial profiles as we did in Sect. 2.3, because
that method did not work for the majority of their cluster sample; the main
reason being that they reached a relatively bright magnitude limit, so that the
low number of members led to uncertainties in the radial profiles. Instead, K13
fitted cumulative radial profiles, with a particular care in choosing the
integration limits and background levels, because of the relative high proportion
of field stars along the cluster line-of-sight \citep{piskunovetal2007}.
Similarly to Fig.~\ref{fig:fig5} (bottom-right panel), they found that clusters in 
common with \citet{froebrichetal2007} show systematically larger core and tidal radii.

\begin{figure*}
\includegraphics[width=\textwidth]{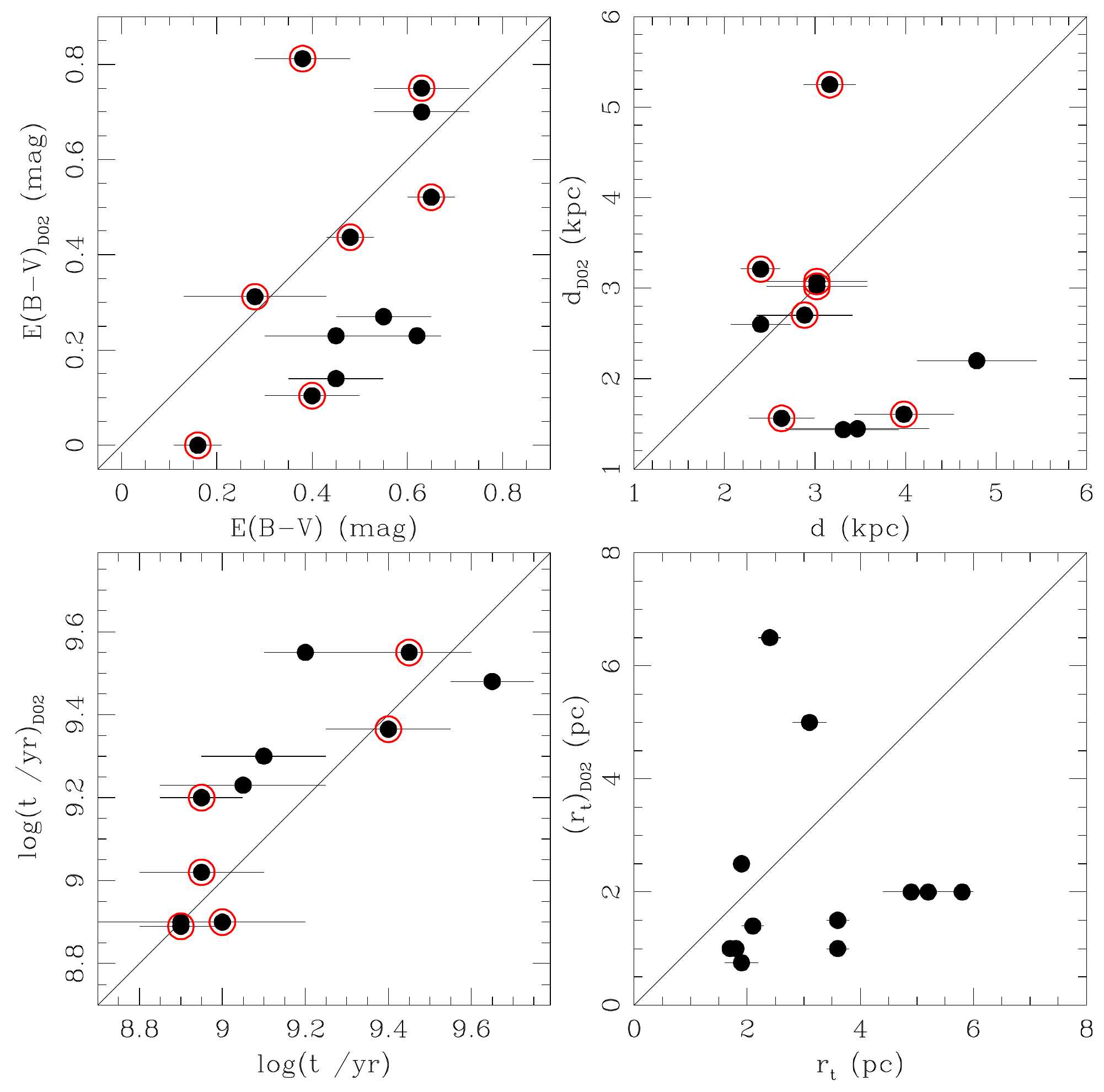}
\caption{Comparison of astrophysical parameters derived by  DAML02
and in this work. Whenever available, error bars are also drawn. The solid line 
represents the identity relationship. Open red circles indicate clusters for which
 DAML02 adopted the parameter values given by K13.}
\label{fig:fig6}
\end{figure*}

DAML02 compiled a catalogue of
open cluster parameters taken from the literature, which we also  compared with the present ones. All the 12 studied clusters are included in  DAML02. 
Fig.~\ref{fig:fig6} illustrates the results.The interstellar reddenings, heliocentric distances
and ages compare with our values similarly as those from K13 did (see Fig.~\ref{fig:fig5}). 
However, the tidal radii -- here we used ( DAML02's diameter)/2 -- are in a better
agreement, thus bringing some additional evidence about the contraints in the K13's
tidal radii mentioned above. In the subsequent analysis, we will use the resulting
parameters derived in this work. 
We notice that the studied clusters are relatively small objects as compared to the 
observed range of sizes of Milky Way open clusters (DAML02).

\begin{figure*}
\includegraphics[width=\textwidth]{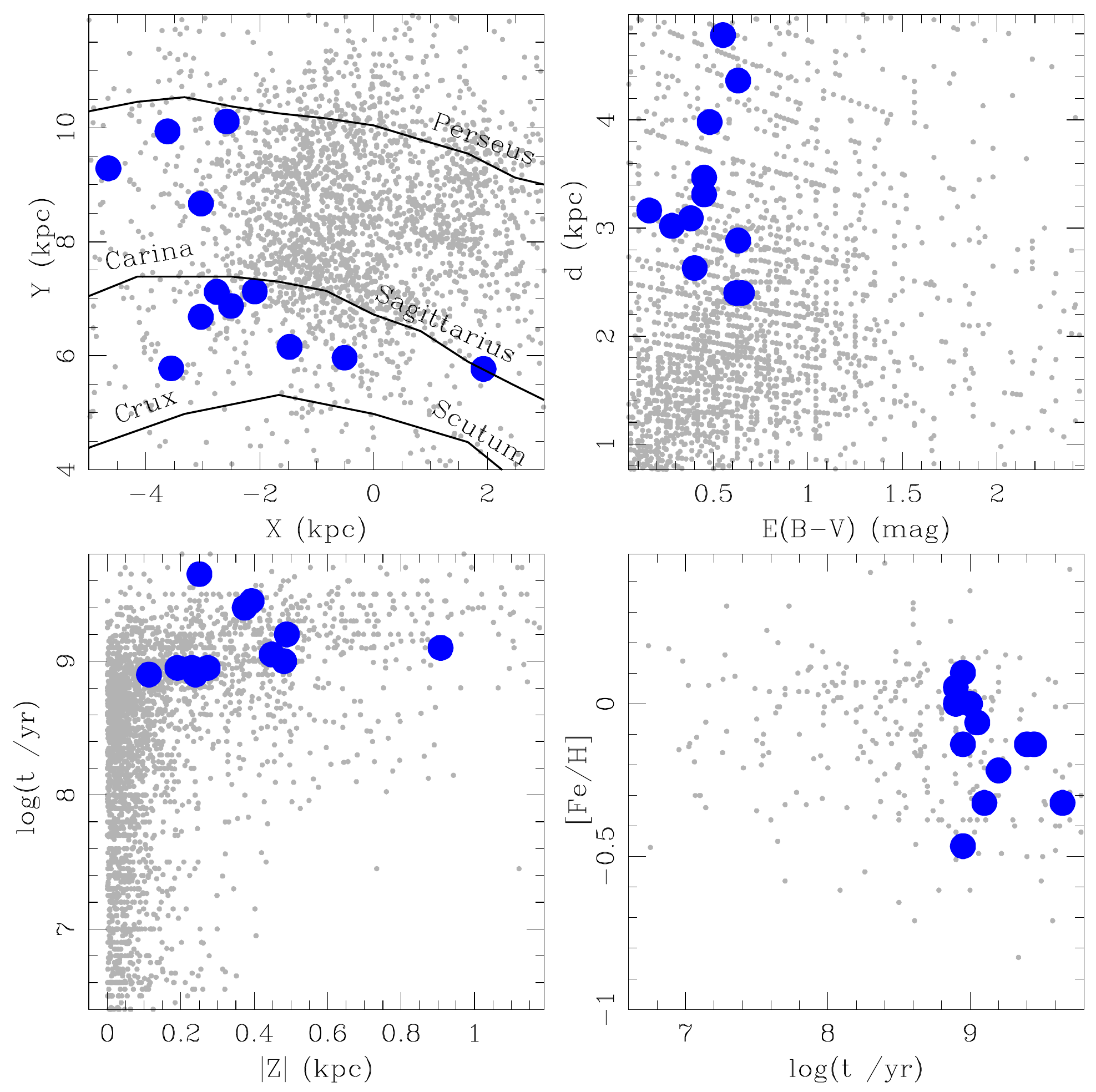}
\caption{Relationships between different clusters' parameters. Large blue and
small grey circles represent the studied and K13's open clusters, respectively,
except in the bottom-right panel, where grey squares are clusters in the
DAML02.}
\label{fig:fig7}
\end{figure*}

Fig.~\ref{fig:fig7} depicts different relationships between cluster parameters with
the aim of bringing the reader an overview of the selected cluster sample. The
top-left panel shows that the studied clusters are located beyond the bulk of
catalogued open clusters. They are not the farthest open clusters detailed studied
until know, but populate a region where studied open clusters are remarkably less in 
number than those distributed within $\sim$ 2 kpc from the Sun, so that the present 
cluster sample objectively contributes to our knowledge to the open cluster system.

Despite their relatively large heliocentric distances, the studied clusters have
interstellar reddenings similar to those located much closer to the Sun (top-right
panel of Fig.~\ref{fig:fig7}), possibly
because they are placed above the Milky way plane ($|$Z$|$ $>$ 0.1 kpc, bottom-left
panel of Fig.~\ref{fig:fig7}). The latter is also a feature of intermediate-age open
clusters that have been formed out of the thick disc's gas, so that due to their orbital
motions they can be found at relatively large heights out of the plane \citep{joshi2018}. 
Interestingly, the clusters' metallicities span a relative wide range in the
age-metallicity relationship (bottom-right panel), in very good
agreement with the observed Milky Way disc metallicty gradient 
\citep{reddyetal2016,magrinietal2017}. Note that we have used ages and [Fe/H] values
from the DAML02's catalogue as a reference. From their present heights, their
intermediate-ages and expected orbital motions, %\citep{vandeputteetal2010}, 
they could have passed across the Milky Way plane several
times, and thus have experienced tidal shocks, interaction with giant molecular clouds, 
etc, that could have contributed somehow in shaping their present internal structures.

It is widely accepted that star clusters lose mass through two main processes,
namely: stellar evolution and disruption caused by the interaction with the 
host galaxy
\citep{lamersetal2005a}. The former is more important during the first hundred Myrs, 
while the latter dominates afterwards, when the mass loss due to stellar evolution 
starts to decrease very slowly \citep{shukirgaliyevetal2018}. Because of the Milky Way 
potential gravitational field varies with the distance from its centre, it is expected
that mass loss by tidal effects follows a similar trend. Hence,
the amount of mass lost in very distance star clusters
should be smaller than those moving in the Milky Way bulge. 

\begin{figure*}
\includegraphics[width=\textwidth]{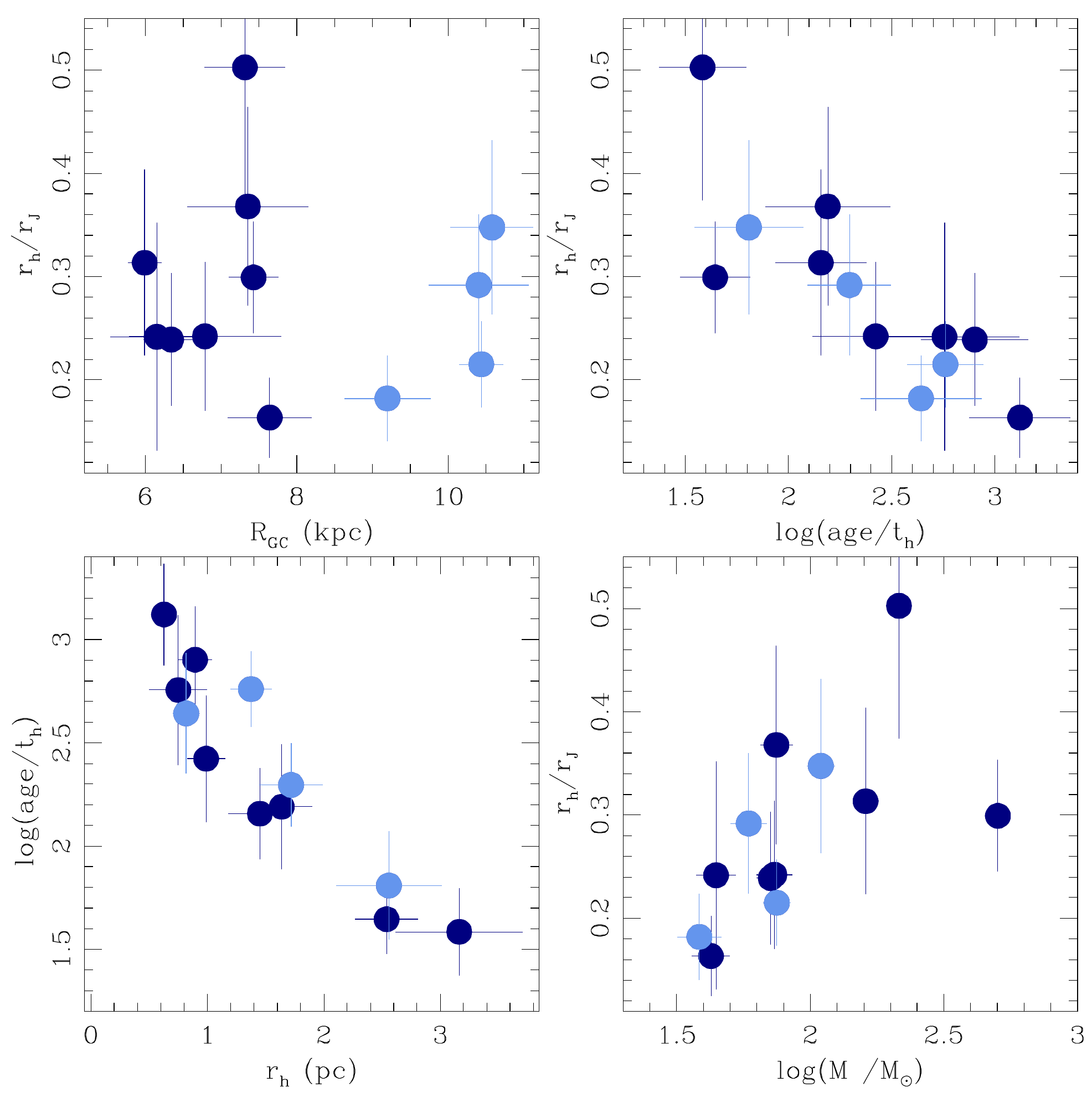}
\caption{Relationships between different structural-dynamics cluster parameters.
Dark-blue and light-blue symbols refer to clusters located inside or outside the
solar circle, respectively. }
\label{fig:fig8}
\end{figure*}

Fig.~\ref{fig:fig7} (top-left
panel) hints at two groups of open clusters among the studied sample, i.e, those located
closer or farther than $\sim$ 8 kpc from the Milky Way centre. In Fig.~\ref{fig:fig8}
(top-left panel) we have drawn the ratio of the half-light radius to Jacobi radius 
as a function of the Galactocentric distance. As can be seen, there is no trend suggesting
that the  $r_h/r_J$ ratios of the clusters in our sample located inside or outside the 
solar circle ($R_{GC_\odot}$ = 8.3 kpc \citet{dgb2017}), which very close to the 
corotation radius \citep[8.51$\pm$0.64 kpc][]{diasetal2019}, have been
differentially affected by the Milky Way tidal field. From this outcome, we interpret
that any difference in the cluster $r_h/r_J$ ratios has been mainly due to
the internal dynamics evolution. Indeed, the top-right panel of Fig.~\ref{fig:fig8}
clearly shows that the $r_h/r_J$ ratio correlates with the times the clusters
have lived their median relaxation times, in the sense that the more dynamically evolved
a cluster, the smaller its $r_h/r_J$ ratio. This means that clusters in more
advanced dynamics evolutionary stages have their integrated light ($\sim$ mass)
more centrally concentrated. Note that these clusters are of nearly
similar intermediate-age, so that the different evolutionary stages would not seem
to come from an age difference but rather from distinct cluster masses. As shown
in the bottom-right panel of Fig.~\ref{fig:fig8}, more massive clusters have
larger $r_h/r_J$ ratios, and hence they are also less dynamically evolved (see 
top-right panel of Fig.~\ref{fig:fig8}). They have also half-light radii larger
than those more dynamically evolved (see bottom-left panel of 
Fig.~\ref{fig:fig8}). 

According to \citet{lg2006}, low mass clusters in the solar neighbourhood are easily 
destroyed by tidal shocks due to e.g. giant molecular clouds and spiral arms. They
found that a cluster located closer than 600 pc from the Sun with an initial mass of 
10$^4$$M_\odot$ is disrupted in $\sim$ 1.7 Gyr, while $\sim$ 0.4 Gyr are needed to
destroy a cluster with a initial mass of 10$^3$$M_\odot$. We cannot directly apply 
these numbers to
the present cluster sample, since they are distributed outside the
solar neighbourhood, but these numbers suggest us that the studied clusters could have 
had initial masses larger than  10$^3$$M_\odot$. Therefore, the small observed masses 
lead us to conclude that these clusters have lost most of their initial masses. Recently, 
\citet{reinacamposetal2019} found in the E-MOSAICS simulations of present-day Milky Way 
mass galaxies \citep{pfefferetal2018} that low-mass clusters lose more mass than those
more massive, in very  good agreement with our findings.

The mechanisms from which the studied clusters could have lost mass are possibly two-body
relaxation following star evaporation. These processes make that the initially dynamically 
warmer inner regions  of a cluster transfer energy to the cooler outer regions, 
so that low-mass stars reach the outer cluster regions, while the cluster core
contracts \citep{portegieszwartetal2010}.  Those low-mass stars that travel over
the Jacobi radius become gravitationally unbounded and disperse into the background
\citep{bonattoetal2004}. Consequently, clusters that have lost more low-mass stars
are relatively more compact (smaller $r_h$ values), and have experienced
more advanced internal dynamics evolutionary stages.

\begin{figure*}
\includegraphics[width=\textwidth]{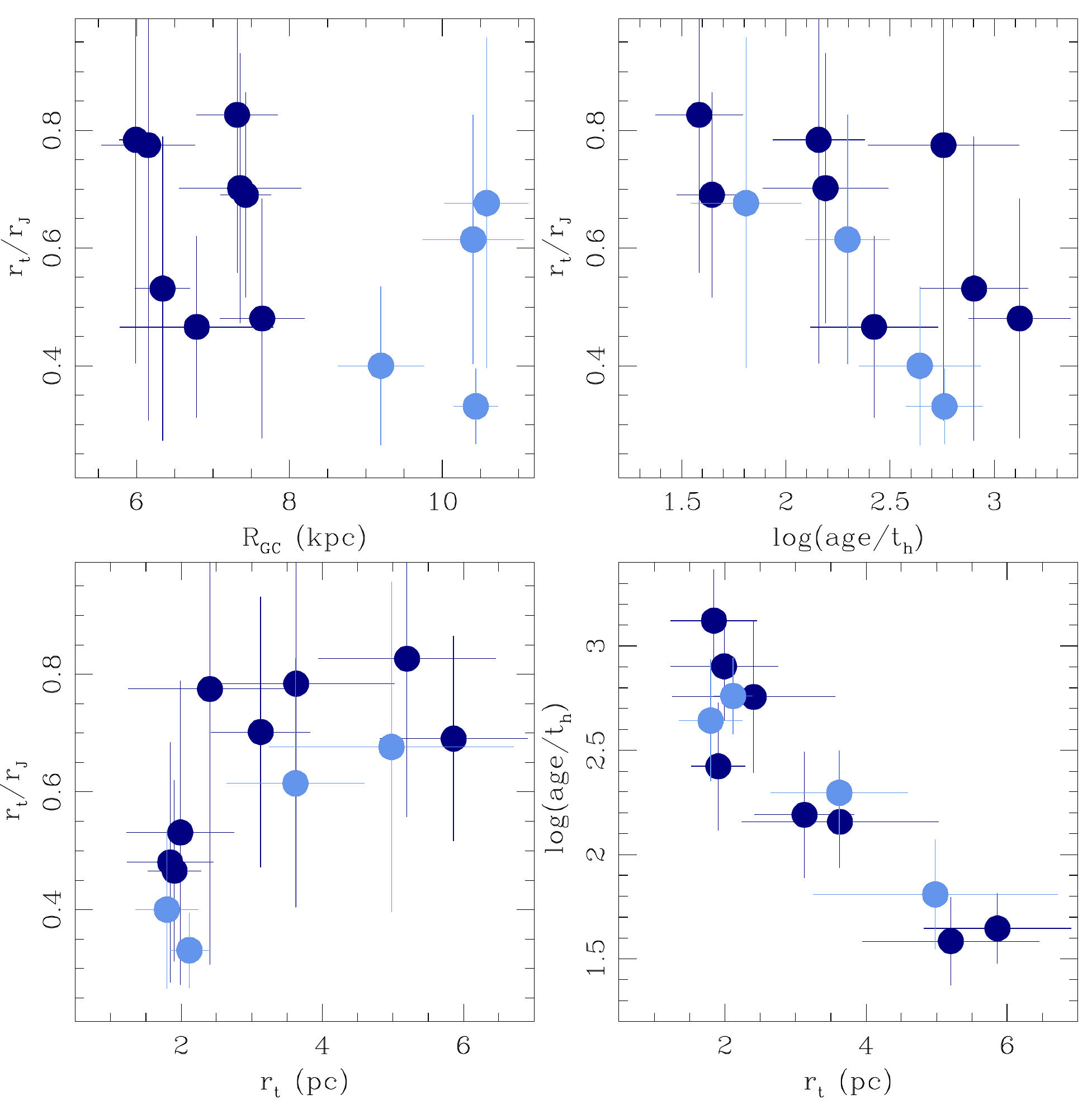}
\caption{Relationships between different structural-dynamics cluster parameters.
Symbols are as in Fig.~\ref{fig:fig8}.}
\label{fig:fig9}
\end{figure*}

\citet{pm2018} showed that the dependence of the gravitational potential with the
distance from the centre of the host galaxy could imprint differential tidal effects
in the outermost structure of the star clusters, while the innermost one were
mostly insensitive to such changes.  We investigated this issue in Fig.~\ref{fig:fig9},
where instead of using half-light radii, we employed tidal ones ($r_t$). The studied clusters
exhibit a very light correlation
with the Galactocentric distance (see top-left panel), in the sense that the
larger the $R_{GC}$ values the smaller the  $r_t/r_J$ ratios. This result could suggest that
the Milky Way gravitational field has been acting differentially on the outermost
cluster regions, making that the expansion due to internal dynamics reached a larger
percentage of the Jacobi volume. 

This combined behaviour of dynamics evolution and tidal effects is also seen in the
top-right panel of Fig.~\ref{fig:fig9}, where the $r_t/r_J$ ratio shows a more
scattered relationship in terms log(age/$t_h$) than that shown from the $r_h/r_J$ ratio
(top-right panel of Fig.~\ref{fig:fig8}). Note that the relationship for clusters located 
inside the solar circle (dark blue symbols in top-right
panel of Fig.~\ref{fig:fig9}) show larger dispersion compared with those of
clusters placed outside the solar circle, possibly because the Milky Way gravitational field
is stronger. Likewise, clusters of similar dimensions ($r_t$) located at 
$R_{GC}$ smaller or larger than $R_{GC_\odot}$ have occupied larger or smaller 
percentages of their Jacobi volumes, respectively. This trend might also be 
attributed to the differential tidal effects. Nonetheless, tidal radii are mainly
driven by internal dynamics evolution (see bottom-right panel of Fig.~\ref{fig:fig9}).

\section{Conclusions}

We have exploited unpublished Washington photometric system images of
intermediate-age open clusters with the aim of characterising their
internal dynamics evolutionary stages. With that purpose, we selected
a sample of 12 open clusters without available deep optical photometry, from
which we derived their astrophysical properties.

The probable cluster members were firstly identified from a proven likelihood
procedure that makes use of  {\it Gaia} DR2 positions, parallaxes and
proper motions. These {\it bonafide} samples of cluster stars were then
used to find the best-solution estimates for the distance modulus, the reddening, 
the age and the metallicity of the open cluster sample. Because of the
well-known degeneracy of these parameters in the cluster CMD, we
applied a method that generated thousands of synthetic CMDs,
taking into account binary effects and covering high resolution ranges in
colour excess, distance, age and metal content. From all of them, we 
employed maximum likelihood statistics and adopted as cluster's parameters
those coming from the theoretical isochrone used to generate the best-matched
synthetic CMD.

The studied open clusters are placed at heliocentric distances of $\sim$
2.4 - 5 kpc and split in two groups, one between the Crux-Scutum and 
Carina-Sagittarius spiral arms and another between the Carina and the
Perseus arms, respectively.  Because of their relative large heights above the 
Milky Way plane (0.1 $<$ $|$Z$|$ (kpc) $<$ 1.0, $<$$|$Z$|$$>$ $\approx$ 0.4 kpc),
they are affected by relatively moderate or low interstellar reddenings
(0.1 $<$ $E(B-V)$ (mag) $<$ 0.7, $<$$E(B-V)$$>$ $\approx$ 0.4 mag). Their
age range spans from $\sim$ 0.8 up to 4.0 Gyr with an average of 1.3 Gyr, and 
they cover the metallicity range [Fe/H] $\approx$ -0.5 - +0.1 dex 
($<$[Fe/H]$>$ $\approx$ -0.1 dex). As far as we are aware, these are the first 
metal abundance estimates derived for these clusters so far.
The resulting astrophysical properties place the 12 studied objects within 
the observed trend of the Milky Way radial and perpendicular metallicity
gradients.
 
From the constructed stellar density radial profiles, corrected from
field star contamination, we derived core, half-light and tidal radii by
minimisation of $\chi^2$ while fitting \citet{king62}'s and \citet{plummer11}'s
profiles. We also obtained their Jacobi radii and half-mass relaxation
times using the above derived parameters and clusters' masses calculated
from mass functions built with stars with membership probabilities
higher than 0.7.

As far as the innermost regions of the clusters are considered, as
traced by the $r_h/r_J$ ratio, we found that their spread of
Galactocentric distances ($\sim$ $R_{GC_\odot}$ $\pm$ 2 kpc) and hence
of the different strength of the Milky Way gravitational field, would
not seem to have had a direct effect in the $r_h/r_J$ ratios. Conversely,
they would rather appear to depend on the stage of internal dynamics evolution
(log(age/$t_h$)), in the sense that the more advanced their internal 
dynamics evolution the smaller the $r_h/r_J$ ratios. Likewise, the
less advanced their internal dynamics evolution, the larger both their
observed masses and half-light radii. These findings reveal that the
more the advanced the internal dynamics, the more compact and less massive
the clusters, a behaviour that place this cluster sample within those
clusters experiencing different levels of two-body relaxation following 
star evaporation.

The outermost cluster regions, monitored by the $r_t/r_J$ ratio with the
Galactocentric distance, show
a slightly different behaviour. Although the internal dynamical clocks
play an important role, the outer cluster regions would seem also
to have been shaped by the Milky Way tidal field. Some evidence of 
such an effect is shown in the subtle dependence of the $r_t/r_J$ ratio
with the Galactocentric distance,
in a more scatted correlation of the $r_t/r_J$ ratios with log(age/$t_h$) 
than that observed for the $r_h/r_t$ ones and in a different percentage of
expansion within the Jacobi volume for clusters located 
inside or outside the corotation radius, respectively. In general, we found that
the farther a cluster from the Milky Way centre, the smaller the
volume occupied within the respective Jacobi one, irrespective of their
actual sizes ($r_t$) and internal dynamics evolutionary stages.

\section*{Acknowledgements}
We thank the referee for the thorough reading of the manuscript and
timely suggestions to improve it.
This work presents results from the European Space Agency (ESA)
space mission Gaia. Gaia data are being processed by the Gaia Data Processing
and Analysis Consortium (DPAC). Funding for the DPAC is provided by national
institutions, in particular the institutions participating in the Gaia
MultiLateral Agreement (MLA). The Gaia mission website is
\url{https://www.cosmos.esa.int/gaia}. The Gaia archive website is
\url{https://archives.esac.esa.int/gaia}.

%%%%%%%%%%%%%%%%%%%%%%%%%%%%%%%%%%%%%%%%%%%%%%%%%%
%%%%%%%%%%%%%%%%%%%% REFERENCES %%%%%%%%%%%%%%%%%%

% The best way to enter references is to use BibTeX:

%\bibliographystyle{mnras}
%\bibliography{paper} % if your bibtex file is called paper.bib

%to be uncommented before sending to editor
%\input{paper.bbl}

% Alternatively you could enter them by hand, like this:
% This method is tedious and prone to error if you have lots of references
%\begin{thebibliography}{99}
%\bibitem[\protect\citeauthoryear{Author}{2012}]{Author2012}
%Author A.~N., 2013, Journal of Improbable Astronomy, 1, 1
%\bibitem[\protect\citeauthoryear{Others}{2013}]{Others2013}
%Others S., 2012, Journal of Interesting Stuff, 17, 198
%\end{thebibliography}

%%%%%%%%%%%%%%%%%%%%%%%%%%%%%%%%%%%%%%%%%%%%%%%%%%

%%%%%%%%%%%%%%%%% APPENDICES %%%%%%%%%%%%%%%%%%%%%

%\appendix

%If you want to present additional material which would interrupt the flow of the main paper,
%it can be placed in an Appendix which appears after the list of references.

%%%%%%%%%%%%%%%%%%%%%%%%%%%%%%%%%%%%%%%%%%%%%%%%%%

% Don't change these lines
\bsp	% typesetting comment
\label{lastpage}
\end{document}